**Novel Double Triple Bend Achromat (DTBA) lattice design for a next generation 3 GeV Synchrotron Light Source**

A. Alekou[1,2], R. Bartolini[1,2], N. Carmignani[3], H. Ghasem[1], S. Liuzzo[3], P. Raimondi[3], T. Pulampong[1], R.P. Walker[1].

[1]Diamond Light Source, Oxfordshire, OX11 0DE, United Kingdom
[2]John Adams Institute, University of Oxford, Oxford OX1 3RH, United Kingdom
and [3]ESRF, 71 Avenue des Martyrs, 38000 Grenoble, France

This project has received funding from the European Union's Horizon 2020 Research and Innovation programme under Grant Agreement No 730871.

**Abstract**
The Double Triple Bend Achromat (DTBA) lattice [1] is a novel lattice design for a next generation 3 GeV Synchrotron Light Source. Starting from a modification of the Hybrid Multi Bend Achromat (HMBA) lattice [2] developed at ESRF and inspired by the Double-Double Bend Achromat (DDBA) lattice [3-4] developed at Diamond, DTBA combines the advantages of both cells. The typical MBA lattice cells have one straight section dedicated to an insertion device, whereas this new cell layout has two such drifts, thus increasing the fraction of available space for the installation of insertion devices and doubling the capacity of the ring. The DTBA lattice achieves an emittance of ~132 pm, in a circumference of 560 m, with a dynamic aperture of ±10 ± 1 mm (calculated at the injection point), an injection efficiency of 88 ±5% and a lifetime of 1.4 ± 0.2 h with errors. The characteristics of DTBA, the methodology and results of the linear and non-linear optics optimisations, with and without the presence of errors, are presented in detail.

*1. Introduction*

Electron storage rings play a very important role in the research of solid state physics, chemistry, cell biology, medicine, protein crystallography and environmental science. They produce high-flux, high-brightness radiation that covers a large spectral region, from far-infrared to hard X-rays. Storage ring-based light sources can easily serve more than 40 experiments simultaneously with excellent reliability, offering a low cost per user. One of the most important measures of the beam quality in storage ring light sources is the brightness, which is the spectral flux per unit volume in transverse phase space:

$$B = \frac{N_\gamma}{4\pi^2 \Sigma_x \Sigma_{x'} \Sigma_y \Sigma_{y'} (\Delta\lambda/\lambda)\Delta t} \quad (1)$$

where $N_\gamma$ is the number of photons within a fractional energy or wavelength band $\Delta\lambda/\lambda$ (typically 0.1%) arriving within time $\Delta t$, and $\Sigma_{x,y}$ and $\Sigma_{x',y'}$ are the root mean square (RMS) size and divergence of the photon beam respectively, in the horizontal and vertical planes.

Hence, the frontier of the next generation light sources is driven by the requirement of an ever decreasing electron emittance reaching the diffraction limit, where the emittance of the electron beam is smaller than that of the photon beam: $\lambda/4\pi$ (8 pm for a wavelength of 1Å). The natural or equilibrium emittance of an electron storage ring is given by:

$$\varepsilon_x = C_q \frac{\gamma^2}{J_x} \frac{\oint H(s)/|\rho|^3 (s)ds}{\oint 1/\rho^2(s)ds} \sim C_q \frac{\gamma^2 \theta^3}{J_x} \propto \frac{\gamma^2}{N_b^3} \quad (2)$$

where $C_q = 3.83 \cdot 10^{-13}$ m is the "quantum constant", $\gamma$ is the Lorentz factor of the electrons, H(s) is the "curly-H" function or dispersion invariant:

$$H(s) = \gamma_x \eta_x^2 + 2\alpha_x \eta_x \eta_{x'} + \beta_x \eta_{x'}^2 \quad (3)$$

ρ(s) is the bending radius of the dipole magnet, θ and $N_b$ are the bending angle and the number of dipoles respectively; $J_x$ is the horizontal damping partition number, with $J_x = 1$ in rings that lack transverse field gradients in the dipoles.

From the second and third part of equation (2) it can be easily seen that in order to obtain a small emittance, a small bending angle or a large number of dipole magnets is needed. Moreover, by using a combined function dipole, $J_x$ will become $J_x>1$, further reducing the emittance. From the first part of (2) it can also be seen that the emittance can benefit from using dipoles with longitudinally varying magnetic field rather than a uniform dipole of a constant bending angle [5-10]. Longitudinally varying dipoles have been studied for the CLIC damping ring [11-12] and are also employed by the ESRF upgrade [2].

The concept of a diffraction limited lattice based on Multi Bend Achromats (MBA) was known since 1993 [13-14], but it was not until the funding of MAX-IV [15] in 2009 that the light source community has employed this approach. The structure of the MBA lattice consists of a set of several unit cells accompanied on each side by a matching section followed by a straight section [13]. The matching sections are necessary to ensure that the dispersion is zero in the straight sections, and that the β-functions meet the requirements of insertion devices (IDs). Table 1 summarises the main parameters and status of (quasi-)diffraction limited rings and upgrades based on the MBA concept.

Table 1: Main parameters of (quasi-)diffraction limited rings and upgrades. The values for Diamond-II refer to the lattice presented in Section 3.3.

|  | E (GeV) | C (m) | $\varepsilon_x$ (pm) | Nat $\xi_x, \xi_y$ | Status |
|---|---|---|---|---|---|
| Diamond-II | 3.0 | 561.0 | 132 | -73,-108 | Concept |
| ESRF-EBS [2] | 6.0 | 844.0 | 140 | -96,-84 | Construction |
| MAX IV [15] | 3.0 | 528.0 | 330 | -50,-50 | Operating |
| SIRIUS [16] | 3.0 | 518.4 | 280 | -113,-80 | Construction |
| SSRF-U [17] | 3.0 | 432.0 | 203 | -74,-59 | Concept |
| APS-U [18] | 6.0 | 1104.0 | 65 | -138,-108 | Pre-construciton |
| ALS-U [19] | 2.0 | 196.5 | 109 | -65,-68 | Concept |
| ELETTRA-U [20] | 2.0 | 260.0 | 280 | -79,-47 | Concept |
| SPRing-8 II [21] | 6.0 | 1320.0 | 150 | -155,-142 | Concept |
| SLS-II [22] | 2.4 | 288.0 | 132 | -69,-34 | Concept |
| HEPS [23] | 6.0 | 1295.6 | 59 | -214,-133 | Concept |

Synchrotron radiation has a significantly higher brilliance (1) when coming from an insertion device (wigglers or undulators) rather than from a bending magnet. Therefore, apart from achieving the smallest possible emittance, another very important parameter for the users is the percentage of the circumference that is free for the installation of the insertion devices. A performance factor is given in [24], defined as the percentage of drift space over the emittance. Table 2 presents this performance factors for a number of light sources. The Double Triple Bend Achromat (DTBA) lattice [1], presented in this paper, gives the best performance factor overall thanks to the small emittance it achieves and due to the fact that it has two times more straight sections compared to the other designs. It provides the options for doubling the capacity of the Diamond storage ring while still providing a small emittance [25].

Table 2: Parameters of 3 GeV storage rings

|  | $\varepsilon_x$ (pm) | $J_x$ | drift % | drift % / $\varepsilon_x$ (nm) |
|---|---|---|---|---|
| MAX IV | 330 | 1.8 | 17.9 | 54 |
| SIRIUS | 280 | 1.3 | 27.1 | 97 |
| Diamond-II | 132 | 1.4 | 31.5 | 239 |
| HMBA [24] | 140 | 1.4 | 25.2 | 179 |
| SSRF-U | 203 | 2.0 | 25.9 | 128 |

## 2. *Optimisation of dynamic aperture and momentum aperture*

The strong focusing required to obtain a small H(s) function, through small $\beta_x$ and $\eta_x$ at dipoles, (see equation (3)), introduces chromatic aberrations which, in order to be corrected, need strong sextupoles. The available stable 6D space (Dynamic Aperture, DA) is limited by the lattice non-linearities, affecting the performance of the ring in terms of injection efficiency, lifetime, etc.

A large optimisation effort is required to improve the injection efficiency of such low emittance lattices, and it is worthwhile to observe that the problem of the DA can be so severe to force the consideration of alternative injection techniques based on the so-called on-axis injection. In fact, the so-called "swap-out" technique has become the baseline design for the planned upgrade of APS [18] and ALS [19] and a wide variety of novel injection schemes has been proposed in the last years [26]. In this paper we aimed at maintaining the classical "off-axis" injection schemes ensuring that the DA, in particular for the projections in the x-y and x-x' planes, is sufficiently large to clear the septum magnet blade (see Figure 1). The introduction of a dedicated injection section with a large $\beta_x$, like the one used for the ESRF upgrade [2], manipulates the 6D phase space volume at injection such that the region where the beam is injected is stable [27]. However, such a cell breaks the symmetry of the lattice and in general reduces the total available DA volume. This approach is studied also for the lattice presented in this paper.

In third generation synchrotron light source rings, the circulating electron bunches have a large charge density and the dominant effect limiting the beam lifetime is Touschek scattering. The key parameter to estimate the beam lifetime in the presence of this effect is the momentum aperture (MA) [28], the energy acceptance at each position in the lattice. The optimization process is based on a suite of different analytical tools that correct the non-linear driving terms, numerical tools, such as genetic algorithms, that focus on the direct optimisation of the DA and MA, and various linear and non-linear optics optimisation strategies are being exploited in order to achieve the best possible storage ring performance. Examples of analytical tools, numerical tools and optimisation strategies, are given in Sections 3 and 4.

The paper is organised as follows: the DTBA layout is presented in section 3. The extensive linear and non-linear optics optimisations followed for the DTBA lattice, is given in section 4. Section 5 presents the effect of the optimisations on DA, Touschek lifetime and emittance in the presence of errors and correction, and finally, the conclusions are given in section 6.

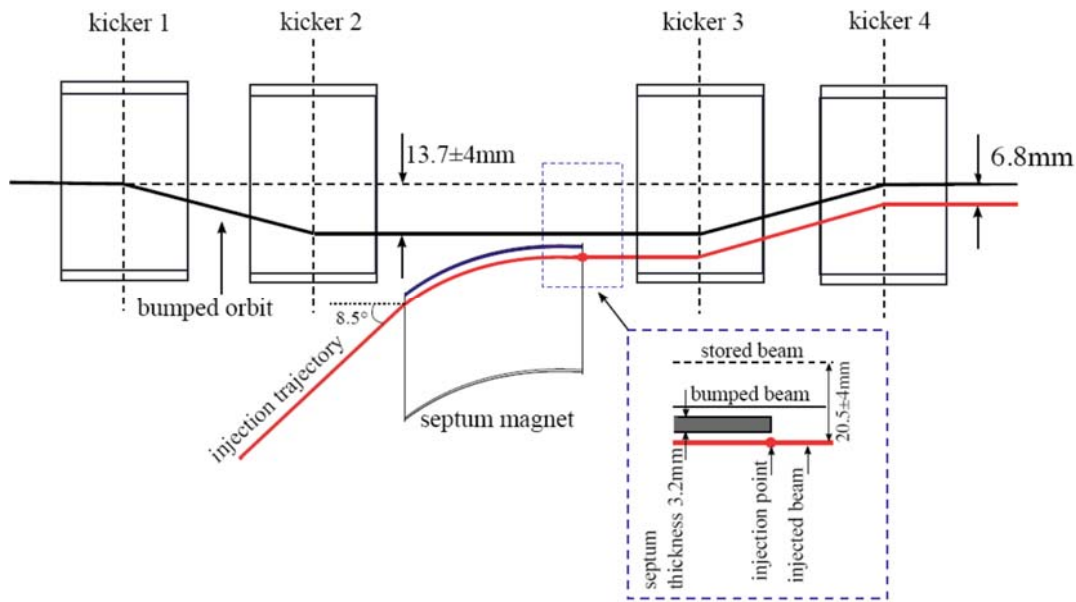

Fig. 1: DTBA off-axis injection scheme

### 3. DTBA layout

As seen in Section 1, the two most important objectives when designing a Synchrotron Light Source are the achievement of high brilliance photon beams, or very small electron beam emittance, and the increase of the number of straight sections, in which insertion devices like wigglers and undulators can be placed. The DTBA lattice was inspired by the DDBA lattice studied for the Diamond Light Source and was created by modifying the ESRF 6 GeV HMBA cell: the central HMBA dipole (DQ2, see Fig. 2) was removed, creating an additional, 3 m long, straight section for insertion devices in the cell centre. The energy was scaled down from 6 GeV to 3 GeV allowing the use of shorter magnets, and the total cell-length was also reduced (see Section 3.2) to match the one of Diamond Light Source. As in HMBA, the DTBA cell (shown in Fig. 3) includes longitudinally varying gradient dipoles (DL), optimised to control the dispersion function and minimize their contribution to the horizontal emittance [29]. The sextupoles are located within two dispersion bumps, which allows them to be more effective and therefore weaker, reducing in this way the geometric aberrations they introduce.

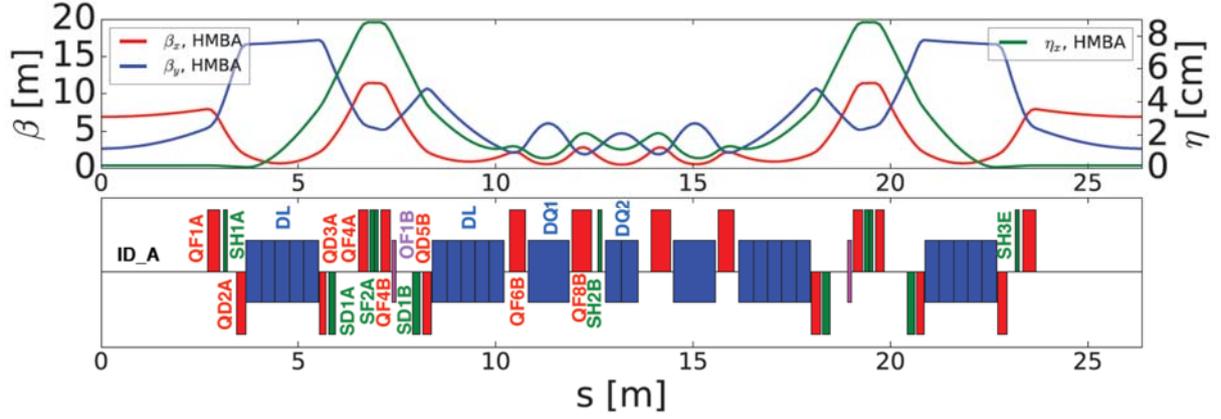

Fig. 2: Optics function of the ESRF-HMBA cell; quadrupoles in red, dipoles in blue, sextupoles in green and octupoles in magenta.

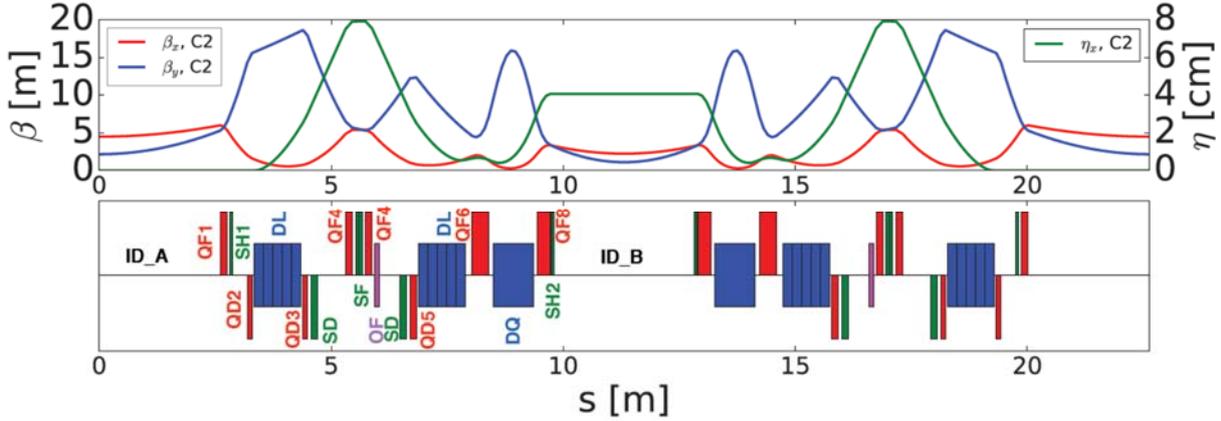

Fig. 3: Optics functions and layout of C2, one of the two main cells for the DTBA lattice; quadrupoles in red, dipoles in blue, sextupoles in green and octupoles in magenta. Detailed optimisation showed that the sextupole SH1 and SH2 can be set to zero. As such, they are not displayed in the following layouts.

In the thin lens approximation the non-linear driving terms will cancel-out if the phase advance between two sextupoles is [30]:

$$\Delta\mu_x = (2n_x + 1)\pi \qquad \Delta\mu_y = n_y\pi \qquad (4)$$

where $n_x$ and $n_y$ are integers. A perfect cancellation of the first order driving terms results in cancellation of the higher order driving terms as the latter are linear combinations of products of the former. As well as in HMBA, this analytical strategy is used in the DTBA cell, where the optics between the two dispersion bumps in a cell are such as to result in a -**I** transformation of phase space between corresponding sextupoles (i.e. between the first and fourth, second and fifth, and third and sixth) [31]. The -**I** transformation can be achieved by tuning the quadrupoles of the cell and the quadrupole field of the combined function dipoles (DQ), which are employed to make the lattice more compact and simultaneously help with the emittance reduction ($J_x > 1$). Due to the presence of interleaved sextupoles the cancellation of geometric aberrations is not exact, and a tuning of the optics and the magnets' strengths becomes mandatory to obtain the required features for the cell with respect to DA and Touschek lifetime (see Section 4). Finally, an octupole family is included as an additional knob to tune the DA.

### 3.1 Linear optics matching knobs

Several optics parameters have to be fixed in the lattice cell in order to achieve the required brilliance, keep the emittance as low as possible and grant the correct phase advances along the cell. This has been summarised experimentally into a set of 9 key optics knobs for 9 quadrupole fields used for the design of the ESRF upgrade HMBA cell [32]. These linear optics knobs allow to determine a unique set of quadrupole strengths and are also approximately linked to main non-linear optics parameters such as detuning with amplitudes and natural chromaticity. These key optics parameters are:

- $\beta_x$ at the ID: the optimal value for undulators is half the drift length, with some flexibility;

- $\eta_x = 0$ at the ID: achromatic condition (this condition can be also replaced by $\eta_x$ at the central sextupoles, SF)
- $\beta_y$, $\alpha_y$ at the central sextupole location (SF): to tune the optics at the available sextupole families;
- the horizontal and vertical phase advance between the two central sextupoles: to grant the **-I** transformation, or tune the deviation from it to an optimal value for beam dynamics;
- the total cell phase advances.

Fig. 4 illustrates the location of these linear optics knobs in the DTBA cell. The knobs have been kept constant when scaling the length of the cell and during the tuning of the quadrupoles' lengths (see Section 3.2). However, once a layout of the correct length and gradients within the limits has been found, those same knobs have been used for the linear and non-linear optimisations as described in Section 3.

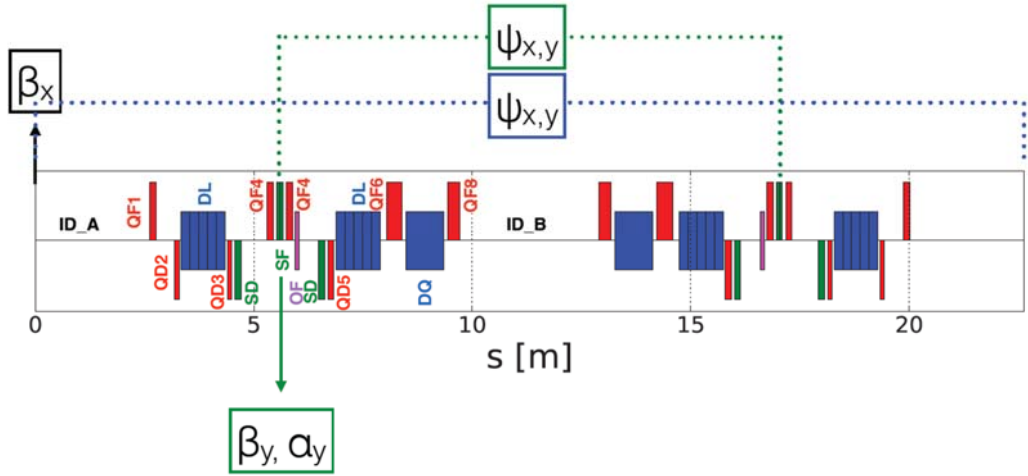

Fig. 4: Linear optics tuning knobs: $\beta_x$ at the ID, $\beta_y$ and $\alpha_y$ at the sextupoles SF, phase advance $\psi_x$ and $\psi_y$ between the two SFs, and total phase advance of the cell.

### 3.2 Modification of the cell length

When designing the DTBA lattice, several geometric constraints of the 3 GeV Diamond Light Source upgrade (Diamond-II) have been taken into account. Diamond-II will have to fit in the tunnel of the current Diamond Light Source so the ID positions are locked points and the total length of 561.0 m can be changed only following a change in the RF harmonic number (within the available RF tuning range). The Diamond lattice consists of 24 cells placed in a 6-fold symmetry; a super-period is composed as (-C1, C2, C2, C1), resulting in a total of 24 straight sections. The DTBA will then have 48 straight sections of three kinds: 18 standard straight sections of 5.1 m length, 6 long straight sections of about 8 m (one of those is the injection straight) and 24 short straight sections of 3.1 m. To design the DTBA cell, C2 has been taken as the basic cell as it is the most frequent and it is symmetric. From C2, C1 and the injection cell can be obtained by performing local modifications. To fit in the lattice layout, C2 must have a length of 22.625 m. The required C2 length was obtained from an initial lattice design derived from the ESRF HMBA, that was 26.4 m long, by reducing the dipoles' and quadrupoles' lengths while keeping all drift lengths fixed. The key optics knobs (see Section 4.1) were kept constant during the process, redefining at each step the new required quadrupoles' strengths. Following an iterative process, the quadrupoles' lengths were also adjusted to ensure all gradients were below 70 T/m. Special care has also been taken for the DQ magnets not to exceed a quadrupole field of 30 T/m and a dipole field of 0.6 T.

### 3.3 Asymmetric DTBA cell

Once the layout of C2 has been fixed, C1 and the injection cell could be derived by modifying only the region of the cell closer to the long straight section. To obtain a layout identical to the one of Diamond, the last drift of C1 should be 1.5 m longer than the last drift space in C2, and should achieve a total straight section length of about 8 m. Although this may seem a simple task, the matching of the C1 cell is not trivial. The key optics knobs were set on C2 and inherited by C1. The additional optics parameters to be set at the long straight section centre (i.e. at the end of C1) were:

- the total phase advance of C1: chosen, as a first approach to keep the lattice symmetry, to be as close as possible to that of C2 (set to $\psi_x = 2.3833$, $\psi_y = 0.8458$);
- $\alpha_x = 0$, $\alpha_y = 0$, $\eta_{x'} = 0$, to match to the next SP;
- the achromatic condition ($\eta_x = 0$) (optional, with the caveats described at the end of the section).

Several solutions have been investigated for the creation of C1:

- C1a: the last C2 drift space was lengthened and one quadrupole was added in the long straight section to allow the matching of the above parameters (Fig. 5 top);
- C1b: an ESRF injection-like cell, created by splitting the last dipole into two uniform dipoles (JL) and placing between them an additional quadrupole (shown in Fig. 5 middle); this design allows to solve simultaneously the C1 and injection cell problems;
- C1c: a cell with a quadrupole triplet in the centre of the long straight section to ease the phase advance and Twiss functions matching (Fig. 5 bottom).

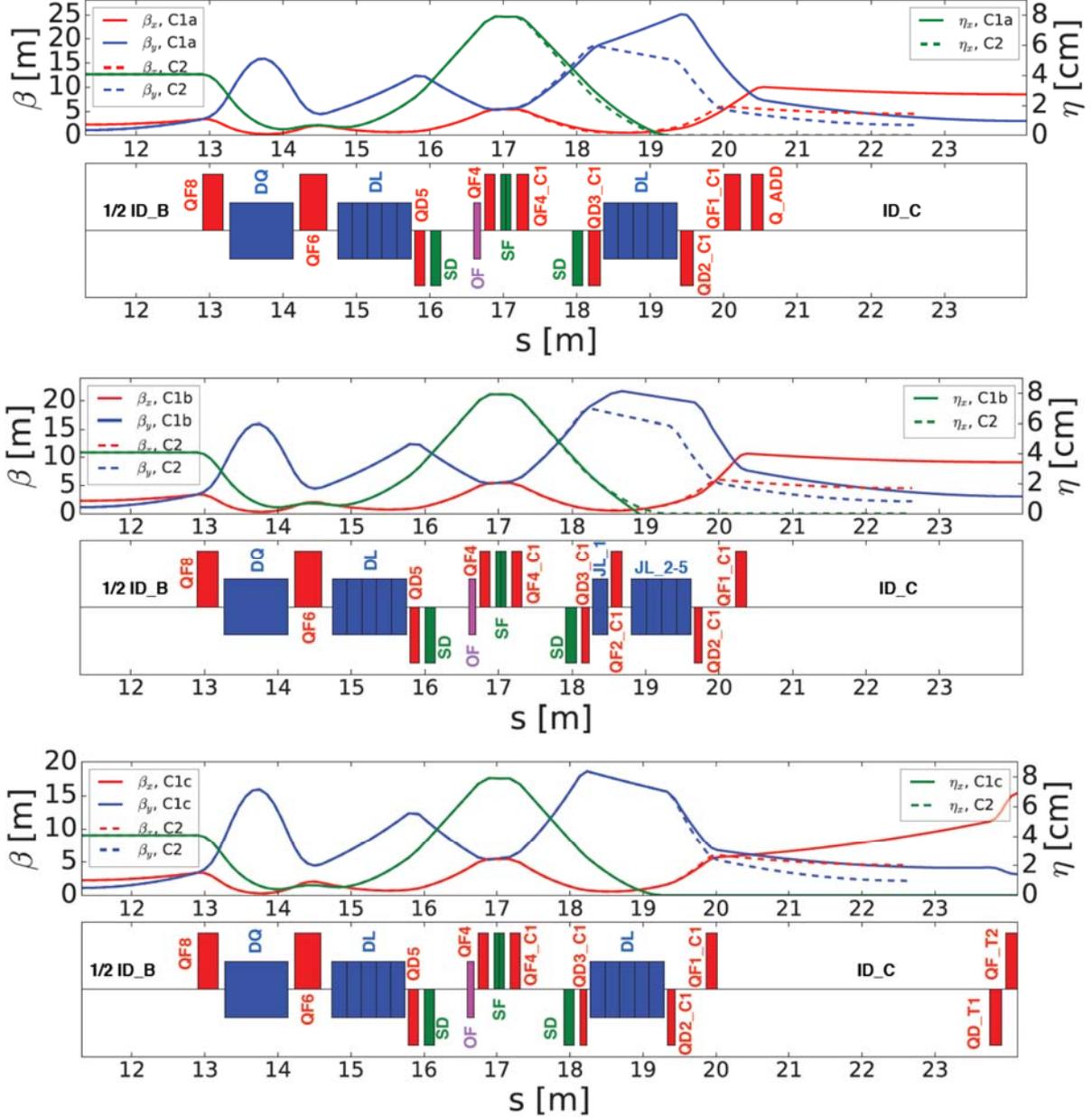

Fig. 5: Summary of optics used for the cell C1: dashed line is the optics in C2 for reference. Cell C1a (top), cell C1b (middle) cell C1c (bottom); quadrupoles in red, dipoles in blue, sextupoles in green and octupoles in magenta.

Fig. 5 shows the layouts of the three C1 designs and a comparison between their Twiss functions (continuous lines) and those of C2 (dashed lines). Notice that in all options the modification of the cell optics is limited to the region after the second focusing sextupole, keeping the first order non linear terms cancellation verified for two over three sextupole families. Also the modification of the optics on the last sextupole is limited and can be partially recovered as described in [33]. Table 3 summarises the main differences between the three C1 versions, and what matching constraints were successfully obtained by each one. For C1a and C1b the last 5 quadrupoles were used for the matching of phase advance and Twiss functions, whereas for C1c the last 6 quadrupoles were used adding a constraint for $\eta_x = 0$. Due to the 70 T/m maximum quadrupole field limitation, the total phase advance of C1a and C1b could not be matched to arbitrary values, and in particular could not be matched to the total phase advance of the symmetric cell C2. The horizontal dispersion at the end of the C1a and C1b cells was ~-4 mm. Note that, although the Long Straight Section (LSS) in C1c is split in two (2x 3.732 m), this was the only version that could be successfully matched to any phase advance and Twiss functions while preserving the achromatic condition. Among these cells, C1a is the simplest but does not allow flexibility in tuning and has the shortest available drift space; C1b and C1c approximate at best the 8 m requirement, however only C1c can be easily tuned. The results obtained for C1b were used for the creation of the injection cell, described in the next Section (3.4) whereas the C1a and C1c versions were used for lattice optimisations, presented in Section 4.

Table 3: Main difference between the three versions of cell C1.

|  | C1a | C1b | C1c |
|---|---|---|---|
| # of quads used for matching | 5 | 5 | 6 |
| Parameters to be matched | $\psi_x = 2.3833$ | $\psi_x = 2.3833$ | $\psi_x = 2.3833$ |
|  | $\psi_y = 0.8458$ | $\psi_y = 0.8458$ | $\psi_y = 0.8458$ |
|  | $\alpha_x = 0$ | $\alpha_x = 0$ | $\alpha_x = 0$ |
|  | $\alpha_y = 0$ | $\alpha_y = 0$ | $\alpha_y = 0$ |
|  | - | - | $\eta_x = 0$ |
|  | $\eta_{x'} = 0$ | $\eta_{x'} = 0$ | $\eta_{x'} = 0$ |
| Ph. adv. match to C2 values | limited range | No | Yes |
| $\eta_x$ at the end of the cell (mm) | -3.96 | -4.33 | 0.00 |
| ID_C (m) | 3.586 | 3.747 | 3.732 |
| Waist at centre of ID_C | Yes | Yes | no |

The characteristics of a DTBA lattice made of 6 SP of (-C1, C2, C2, C1) are presented in Table 4 and the layout of one SP is shown in Fig. 6.

Table 4: Characteristics of the DTBA built with six superperiods named SP made of (-C1, C2, C2, C1).

|  | -C1a, C2, C2, C1a | -C1c, C2, C2, C1c |
|---|---|---|
| Circumference (m) | 560.0 | |
| C1, C2 [m] | 24.125, 22.625 | |
| ID_A, B [m] | 2.606, 3.180 | |
| $v_x, v_y$ | 57.20, 20.30 | |
| nat. $\xi_x, \xi_y$ | -73.05, -105.72 | -73.10, -108.01 |
| $\varepsilon_x$ (pm) | 131.24 | 131.45 |
| $\alpha_c$ [$10^{-4}$] | 0.99 | |
| $U_0$ [MeV] | 0.36 | |
| $\sigma_z$ [mm] | 1.71 | |
| RF voltage [MV] | 2.2 | |
| RF frequency [MHz] | 499.65 | |
| Harmonic number | 935 | |

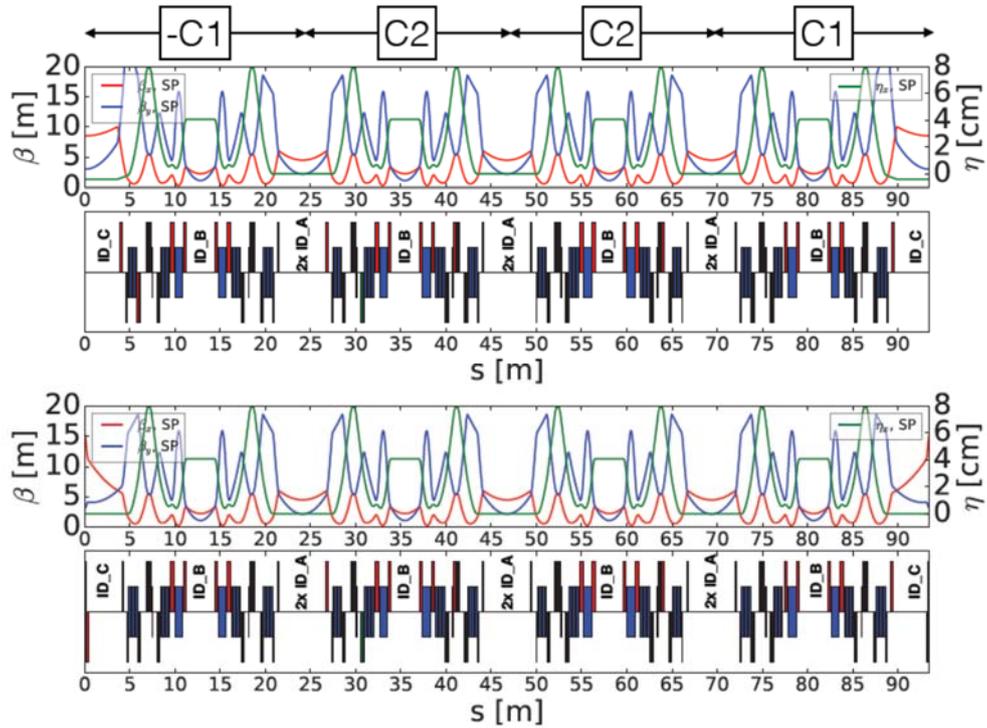

Fig. 6: Twiss functions and layout of the whole superperiod SP (-C1, C2, C2, C1) using C1a (top) and C1c (bottom). The three different ID straights are indicated.

In the DTBA lattice further care must be taken in assessing the impact of the Insertions Devices on the emittance in particular if strong IDs will be installed in the mid-straight section which has a non-zero dispersion. The emittance growth as a function of the on axis ID field is reported in Fig. 7 and shows that a strong ID cannot be installed unless the dispersion is better controlled. Likewise a reduction in the dispersion will also produce a reduction in the effective emittance for the source in the mid-straight section. These aspects however can be tackled with a further optimisation of the lattice functions.

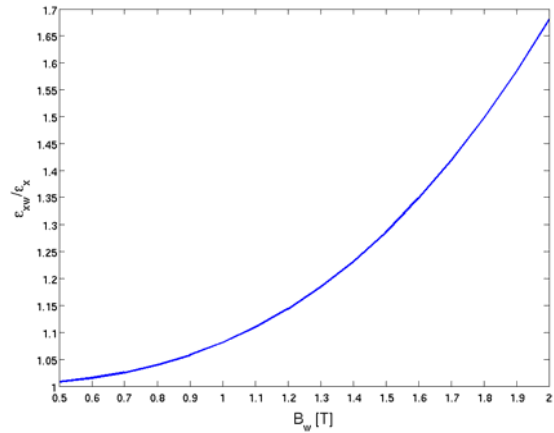

Fig. 7: Emittance growth as a function of the ID strength for different values of peak field on axis of the ID in the mid-straight section.

### 3.3 Injection cell
As mentioned in the introduction, several techniques are being employed to increase the injection efficiency, which is reduced due to the non-linearities of the lattice. In the Diamond-II case the injection chain is already defined and the beam will be injected off-axis at about -6.8 mm from the stored beam in the radial direction. The design of a dedicated injection section with large horizontal beta function allows to adjust the phase space at the injection location and grant a stable region to store the injected beam.

Based on the ESRF injection cell [34], the DTBA injection cell, CINJ, was created by splitting the last longitudinal-gradient dipole in two uniform dipoles, in order to enable the matching of dispersion to small values. A notable feature of this cell is that, without changing the total cell phase advance, the desired $\beta_x$ can be matched at the injection point only by altering the location of the last quadrupole of C1b, QF1-I (see Fig. 5b). Three different injection cells with $\beta_x = 20$ m, $\beta_x = 30$ m and $\beta_x = 45$ m at the injection straight, were thus simply obtained by moving QF1-I and matching the optics (see Fig. 8). No limits were set for the quadrupoles of the injection cell, however all of them were below 70 T/m.

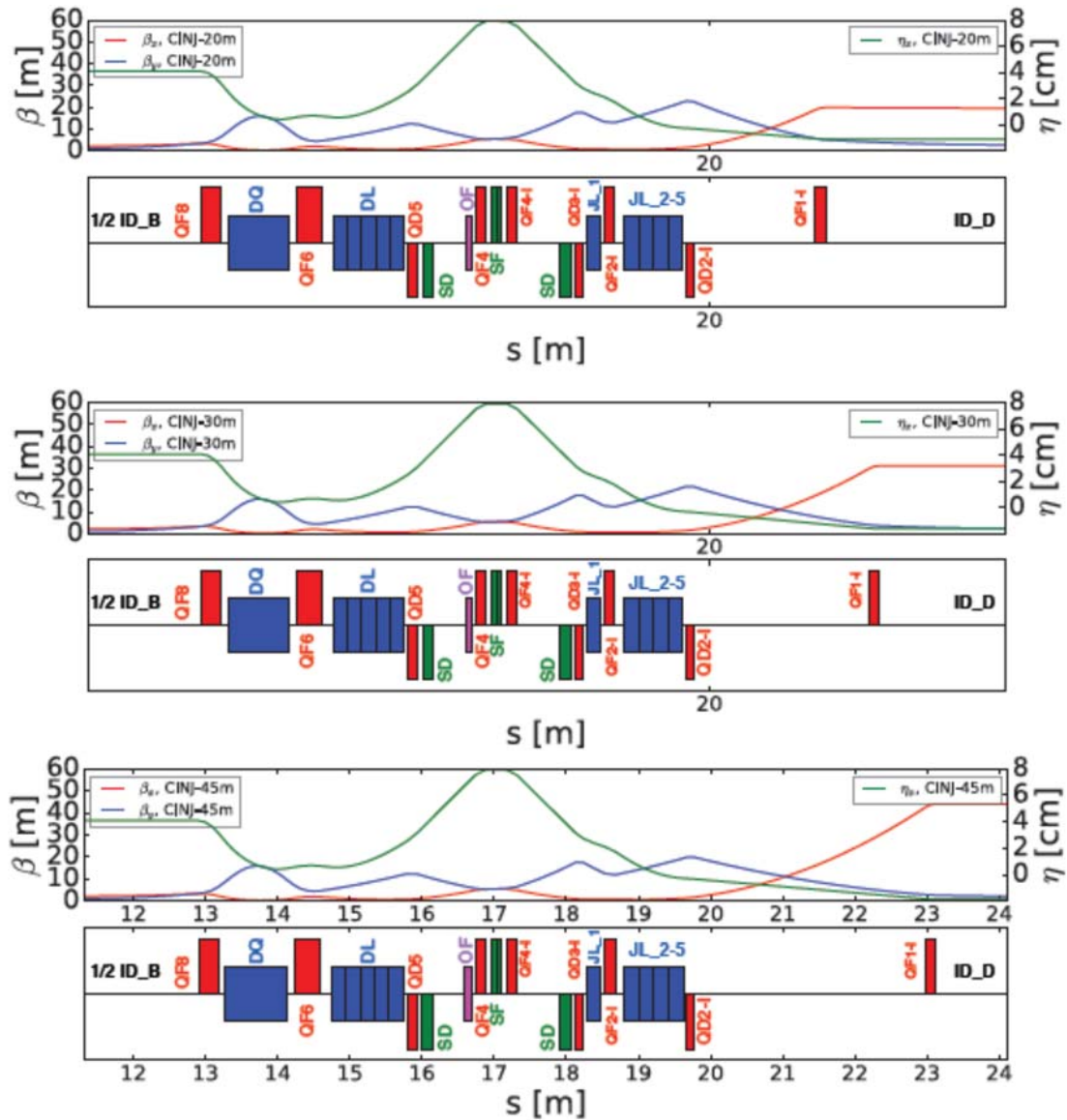

Fig. 8: Twiss functions and layout of the injection cell CINJ, from the middle to the end of the cell, for design with the $\beta_x = 20$ m (top), $\beta_x = 30$ m (middle), $\beta_x = 45$ m (bottom); quadrupoles in red, dipoles in blue, sextupoles in green and octupoles in magenta. Note how the value of $\beta_x$ depends on the position of the last quadrupole QF1-I.

The DA of each of the three $\beta_x$ designs, calculated at the centre of ID_A and at the centre of the injection cell straight (ID_D), is presented in Fig. 9 (top), with continuous and dashed lines respectively. The DA of a DTBA ring without an injection cell, calculated at the centre of ID_A, is also plotted (cyan continuous line). Comparing the periodic solution (cyan line) to the lattices including the injection cell (continuous lines) it is clear that the DA, with no errors, is not affected negatively in the presence of the injection cell. The best DA at injection is the one that corresponds to $\beta_x = 45$ m (blue dashed line). On the other hand, the MA has been reduced significantly when the injection cell has been included. Fig. 9 (bottom) compares the MA of DTBA with (continuous line) and without (cyan dashed line) the injection cell. The injection efficiency was found to be 79.1%, 78.6% and

75.8% for the lattice with $\beta_x = 20$ m, $\beta_x = 30$ m and $\beta_x = 45$ m at injection respectively. Since the best MA and injection efficiency is obtained with the $\beta_x = 20$ m lattice (red line), which also has an acceptable DA with no errors, this version was selected to be used in the optimisation described in Section 4.

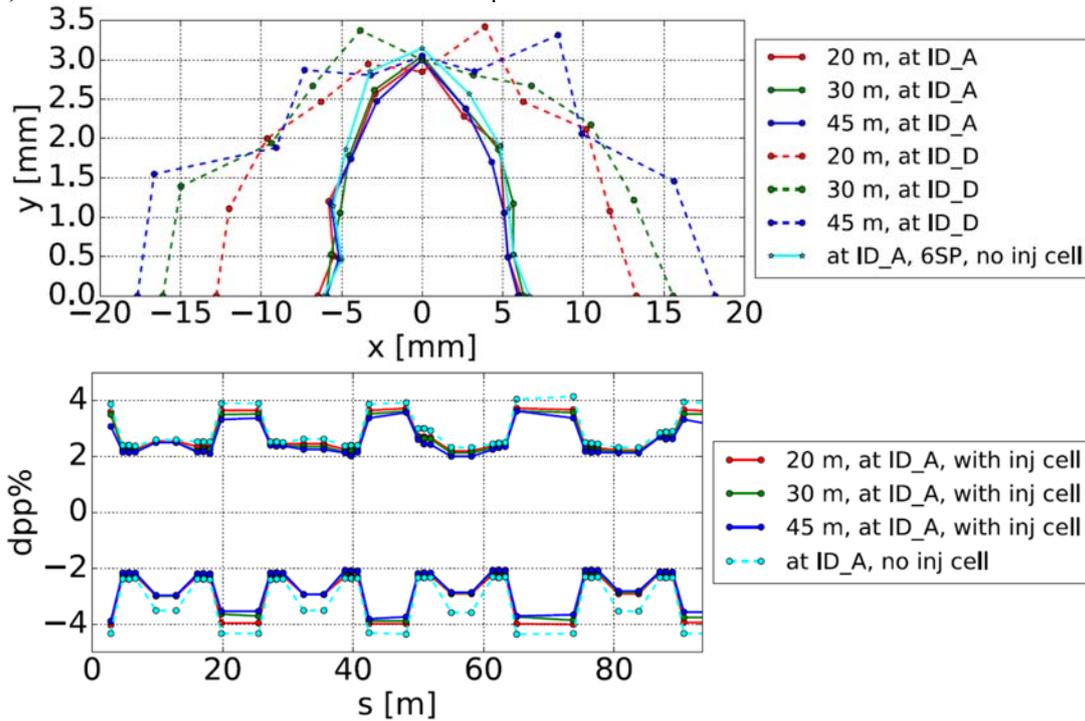

Fig. 9: DTBA DA (top) and MA (bottom) without errors, with and without the injection cell

The first 100 m of a full DTBA ring, including the injection cell with $\beta_x = 20$ m, is shown in Fig. 10; the natural chromaticity and emittance are presented in Table 5.

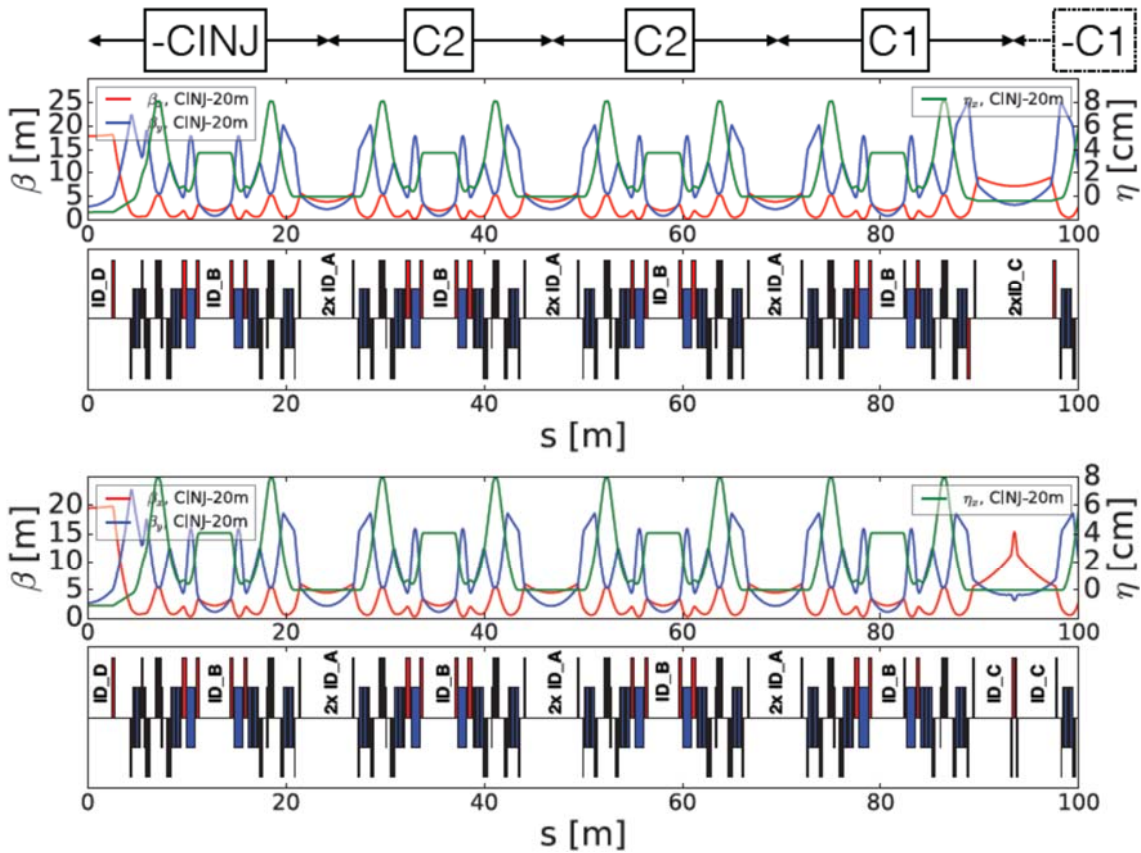

Fig. 10: Twiss functions and layout of the first 100 m of the DTBA ring with $\beta_x = 20$ m at injection, using C1a (bottom) and using C1C (top); quadrupoles in red, dipoles in blue, sextupoles in green and octupoles in magenta.

Table 5: Natural chromaticity and emittance of the DTBA lattice including the injection cell.

| C1 version | C1a | C1b |
|---|---|---|
| nat. $\xi_x$, $\xi_y$ | -70.54, -116.09 | -70.38, -92.29 |
| $\varepsilon_x$ (pm) | 130.35 | 124.76 |

## *4. DTBA lattice optimisation*

Several iterations between the linear and non-linear optics optimisations are essential for the maximisation of the lattice performance with respect to DA and MA. A good starting point for the linear optics optimisation is the choice of a working point that avoids destructive resonances. The choice of minimum and maximum Twiss β-functions are important as they contribute to the magnitude of the natural chromaticity, and the higher the dispersion the lower the required sextupole strengths. Although a resonance cancellation scheme can be beneficial (see Section 3) in providing a good starting point, advanced numerical optimization techniques provide usually further improvement on the non-linear optics optimisations. A combined effort in the tuning of optics functions, sextupoles and octupoles, and phase advances, is crucial to maximise the DA and MA.

The combination of large computer clusters and parallelised algorithms play a very important role in the optimisation of ultra-low emittance rings. The DA and MA of the ring can be optimised using calculations that include full 6D tracking with realistic machine models including random errors, physical apertures, insertion devices, etc [35]. As described in [36], a systematic scan of quadrupoles and sextupoles (e.g. Global Linear Analysis of All Stable Solutions (GLASS)) is useful in rings with a limited number of magnet families. In the case of more than five dimensions different methods should be explored. Multi-Objective Genetic Algorithms (MOGA), a highly effective randomly searching algorithm based on Non-dominated Sorting Genetic Algorithm (NSGA-II) [37-38], can search a set of solutions over a domain of many dimensions with multiple conflicting objectives [39]. When considering all of the objectives, the solutions that are superior to the rest are called Pareto-optimal solutions, and among those the user can choose the best compromise. This algorithm in conjunction with storage ring lattice design codes such as MAD [40], Elegant [41] etc, for single particle 6D tracking, can help spare much of the designer's time and enable a direct judgment of different solutions in terms of their ultimate simulated performances (DA, Touschek lifetime). MOGA has been proven to be a suitable tool for high computational complexity problems and in particular it has been successfully used in the past to perform linear and non-linear optimisations of particle accelerator lattices [42-44]. For the above reasons, MOGA is used for the optimisations presented in this paper. The single particle tracking codes Elegant [41] and Accelerator Toolbox (AT) [45] were used for e- tracking. Note that no physical apertures or insertion devices were included in the optimisations. A final check of the beam dynamics with misalignments and magnetic field errors is carried out as a final benchmark of the robustness of the optimisation.

## *4.1 Linear optics optimisation*

Each of the linear optics knobs defined in Section 3.1 is also a key knob that influences (mainly) a well defined quantity of interest (see Fig. 11 and Table 6). For example, Fig. 12 shows the effect of $\beta_x$ at ID_A on emittance, DA (calculated at ID_A) and Touschek lifetime (for the purposes of this example the $\beta_x$ scan was performed on a ring made of 24 C2, while keeping all other knobs of Table 6, and the phase advance of the cell, constant).

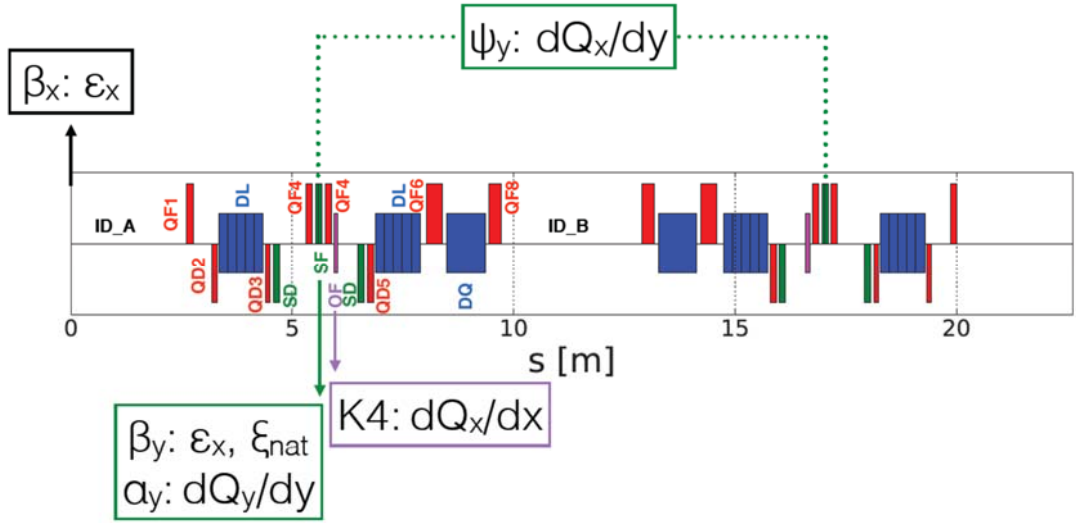

Fig. 11: Layout of C2 with the optics tuning knobs and the corresponding dynamical quantities they influence.

Table 6: Optics tuning knobs inherited from the optimisation of the HMBA lattice

| Parameter | Location | Influence |
|---|---|---|
| $\alpha_y$ | SF | V detuning with V amplitude |
| K4 | OF | H detuning with H amplitude |
| $\psi_y$ | SF-SF | Cross detuning term |
| $\beta_x$ | ID_A | $\varepsilon_x$ |
| $\beta_y$ | SF | $\varepsilon_x$ and natural chromaticity |

For a complete optimisation, the parameters of Table 6 have been varied simultaneously using MOGA, by altering all available quadrupole components, with lifetime and x-y dynamic aperture as objectives. During the optimisation process the total phase advance of the cell was kept constant and the chromaticity was matched to (2, 2) [46]. The lifetime calculations were performed using a current of 300 mA, 900 bunches, bunch-length $\sigma_z$ (zero-current) = 1.71 mm and 10% coupling. The results of the Pareto front of this optimisation are shown in Fig. 13 for a DTBA lattice using the C1a (left) and C1c (right) versions. Note that at this stage of the optimisation the injection cell was not included. Also, the increase of Touschek lifetime was a more important objective than DA, as it was expected that with the inclusion of the dedicated insertion cell, described in Section 4.4, we would be able to remodel the phase space at the injection point to inject in a stable region.

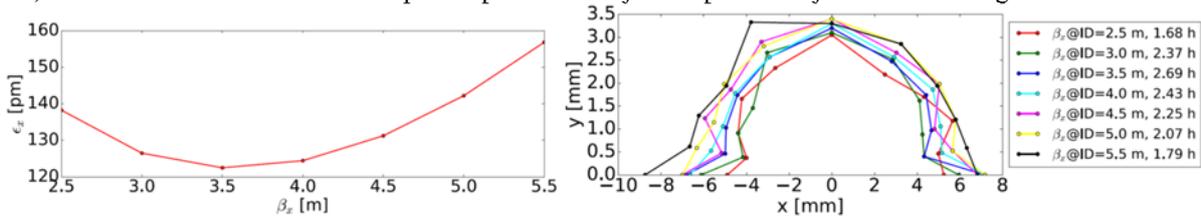

Fig. 12: Effect of $\beta_x$ at ID_A and on $\varepsilon_x$ (left) and on DA and Touschek lifetime (right).

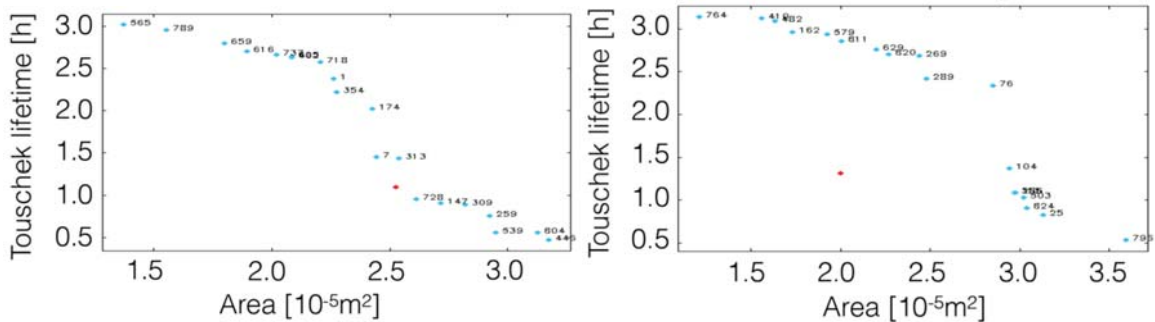

Fig. 13: Optimisation of the linear optics knobs of a DTBA lattice with C1a (left) and C1c (right) using MOGA. Red dot: initial point, blue dots are the Pareto front with the best solutions found so far.

The linear optimisation had a very positive impact on lifetime which was increased from ~1 h to up to ~3 h. Note that for τ ~ 2.5h the DA is better for the C1c solution than for C1a. It should be stressed that throughout these optimisations the linear optics knobs are tuned to achieve optimal non-linear lattice parameters (DA and Touschek lifetime). This is a remarkable feature of this kind of lattices, that highlights the necessity to tune simultaneously linear and non-linear parameters from the early stages of the lattice design.

### 4.2 Non-linear optics optimisation

Following the linear optics knobs optimisation, the sextupoles and octupoles of the DTBA lattice have been optimised using MOGA. The sextupoles of the SP were divided in 6 families: SD1, SF1, SD2 in C2, and SD3, SF2 and SD4 in C1. C1 and C2 had an octupole family each. The chromaticity was matched to (2, 2) using SD1 and SF1, whereas SF2, SD3, SD4, OF1 and OF2 were used for the optimisation of the DA area and Touschek lifetime. Fig. 14 presents an example of the first member of a generation (red) and the Pareto front of the best found solutions (blue), for a DTBA with C1a (left) and C1c (right) solutions. The non-linear optics optimisation showed a stronger improvement in DA rather than the Touschek lifetime. Note that no errors have been included in this optimisation, due to the early stage of the lattice design. However, it is important to stress that a remarkable feature of the MOGA-type of optimisation is that it can be applied to lattices in presence of errors and this remains true in the online optimisation of existing machines [47].

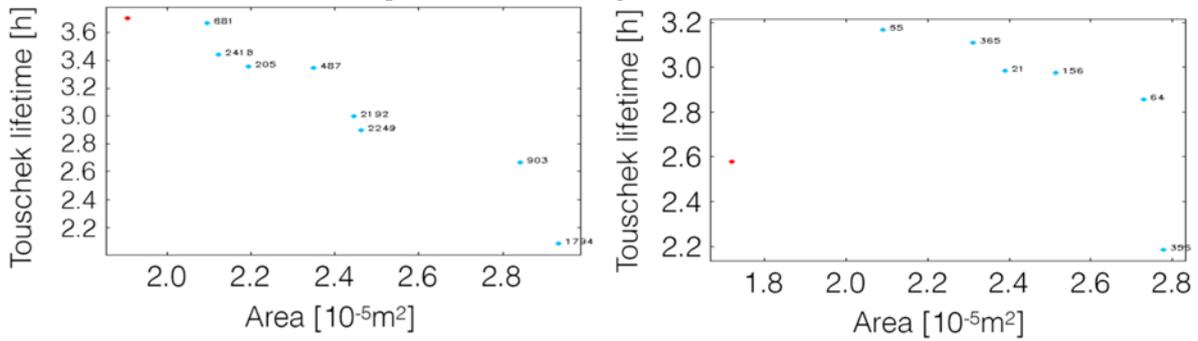

Fig. 14: Sextupole and octupole optimisation of the DTBA lattice using MOGA for the DTBA lattice with C1a (left) and C1c (right). Red dots: first member of a generation, blue dots: the Pareto optimal front.

### 4.2 Injection efficiency optimisation

After a satisfactory solution was found, a DTBA lattice including the injection cell was optimised using the procedure described above. The non-linear optimisation included additional sextupole and octupole families for the injection cell: SF3, SD5, SD6 and OF3. The objectives of this optimisation changed from DA and lifetime, used in the linear and non-linear optics optimisations, to injection efficiency and lifetime. The Pareto front of the optimisation is shown in Fig. 15 (left: with C1a, right: with C1c). The injection efficiency reaches ~85%, however the lifetime was found to be about 1h. Since the optimisations of both C1a and C1c led to similar results in terms of DA, Touschek lifetime and injection efficiency, and C1c can be tuned to any optics and has a smaller natural chromaticity than C1a (see Table 5), we decided to pursue the studies using C1c and left C1a for future investigations.

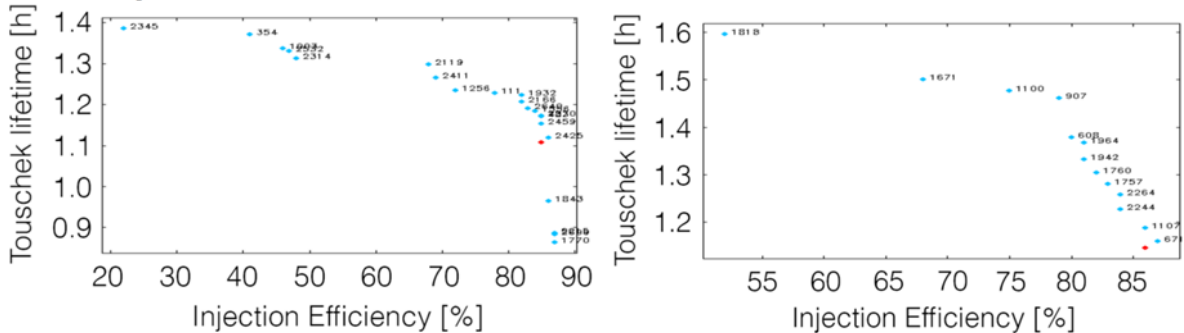

Fig. 15: injection efficiency and Touschek lifetime with linear and non-linear optics optimization (no errors), including the injection cell, using MOGA for the DTBA with C1a (left) and C1c (right). Red dots are the first members of a generation, blue is the Pareto optimal front.

### 4.3 Empirical optimisation

A semi-empirical approach has been also adopted to optimise the sextupoles, octupoles and phase advance (total and between sextupoles) for the three different cells, C1c, C2 and CINJ. For each knob of interest the DA, emittance and lifetime have been computed for several different values and a choice has been made among the tested values in order to obtain an optimal compromise. Following an iterative process the DA and lifetime of the DTBA cell were further improved, achieving about 12 mm DA at the injection point and 2.3h lifetime without errors. The effect of the SH2 sextupoles at the extremities of the central straight section has also been considered. As shown for the Spring-8 upgrade [48], a small field in those magnets helps improve the DA and lifetime. During this process the possibility to break the requirement of constant cell phase advance was tested. This did not seem to be of any advantage; however only a limited range of variation has been explored. The final optimised magnets strengths are listed in Table 7.

Table 7: DTBA optimized parameters

| Magnet name | L (m) | $B_0$ (T) | $b_2$ (T/m) | $b_3$ (T/m$^2$) |
|---|---|---|---|---|
| DL1A_1 | 0.201 | 0.255 | | |
| DL1A_2 | 0.201 | 0.294 | | |
| DL1A_3 | 0.201 | 0.344 | | |
| DL1A_4 | 0.201 | 0.425 | | |
| DL1A_5 | 0.201 | 0.650 | | |
| DQ | 0.867 | 0.598 | -28.3 | |
| JL | 0.204 | 0.389 | | |
| QD2 | 0.102 | | -64.5 | |
| QD2-I | 0.102 | | -51.3 | |
| QD2_C1 | 0.102 | | -53.1 | |
| QD3 | 0.100 | | -57.6 | |
| QD3-I | 0.150 | | -63.7 | |
| QD3_C1 | 0.100 | | -56.7 | |
| QD5 | 0.136 | | -44.3 | |
| QD_T1 | 0.162 | | -40.6 | |
| QF1 | 0.155 | | 62.3 | |
| QF1-I | 0.100 | | 36.7 | |
| QF1_C1 | 0.155 | | 49.3 | |
| QF2-I | 0.150 | | 59.4 | |
| QF4 | 0.142 | | 48.5 | |
| QF4-I | 0.142 | | 45.9 | |
| QF4_C1 | 0.142 | | 48.4 | |
| QF6 | 0.362 | | 64.7 | |
| QF8 | 0.278 | | 64.1 | |
| QF_T2 | 0.162 | | 41.6 | |
| SD1 | 0.140 | | | -1919.3 |
| SD2 | 0.140 | | | -2285.0 |
| SD3 | 0.140 | | | -1919.3 |
| SDJ | 0.140 | | | -1919.3 |
| SF1 | 0.140 | | | 2234.9 |
| SH1A | 0.050 | | | |
| SH2B | 0.050 | | | 100.1 |
| SH2D | 0.050 | | | 100.1 |
| SH3E | 0.050 | | | |

*4.4 Errors and corrections*
Following the optimisations described in Section 3 and 4, the performance of the DTBA lattice with C1c was studied in the presence of errors and correction. Random alignment and roll errors were set on the magnets according to the values listed in Table 8. Orbit, optics and coupling were corrected following [49] and the DA (computed at the ID_A centre) and lifetimes were estimated for 50 seeds of errors. The resulting DA computed at the injection point and lifetime, averaged over 50 seeds, were found to be 9.8 ± 1.1 mm, which is sufficient for injection at 6.8 mm, and τ = 1.4 ± 0.2 h. Fig. 16 shows the effect that the errors have on the DA. To achieve these values several BPMs have been distributed in the lattice to correctly sample the orbit and optics functions. Further optimisations will aim to reduce the number of required BPMs. Table 9 lists a series of relevant quantities before and after correction with the available independent power supplies.

Table 8: Random errors: rms of Gaussian distributions truncated at 2.5 σ

| Magnet | Δx (μm) | Δy (μm) | Roll angle (μrad) | Field integral error |
|---|---|---|---|---|

| Dipoles (DL) | 50 | 50 | 50 | $10^{-4}$ |
| DQ | 50 | 50 | 50 | $10^{-4}$ |
| Quadrupoles | 50 | 50 | 50 | $5 \cdot 10^{-4}$ |
| Sextupoles | 50 | 50 | 50 | $35 \cdot 10^{-4}$ |
| BPMs | 50 | 50 | | |

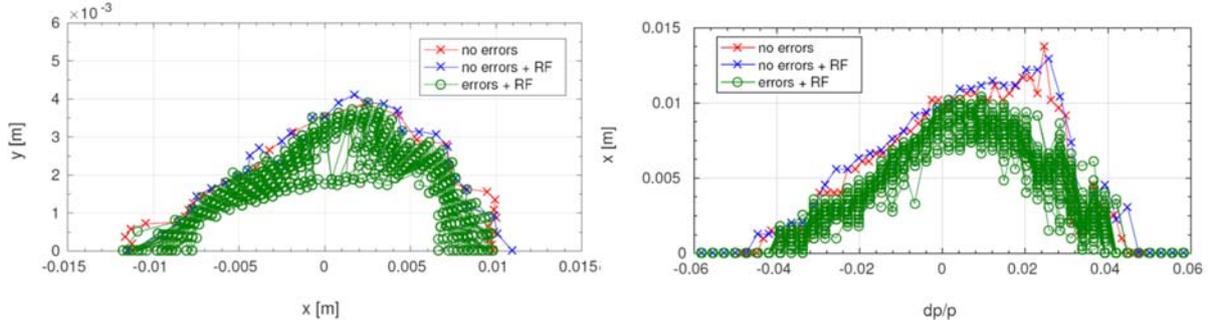

Fig. 16: DA on-energy (left) and off-energy (right) of the DTBA lattice with C1c with no errors and no RF cavity (red), no errors and with the RF cavity (blue), with errors and RF cavity (green).

Table 9: Average over 50 seeds of errors after correction of orbit, β-beating, dispersion tunes and emittance

| | X | Y |
|---|---|---|
| orbit (μm) | 79 | 62 |
| Δη (mm) | 0.2 | 0.4 |
| Δβ/β (%) | 0.6 | 0.6 |
| $\xi_x, \xi_y$ | 2 | 2 |
| $\nu_x, \nu_y$ | 57.195 | 20.383 |
| $\varepsilon_x$ (pm) | 132 | 0.05 |
| steerers (mrad) | 0.14 | 0.24 |

The injection efficiency estimations for the 50 seeds studied lead to 88 ± 5% injection efficiency. These values were computed using a beam with $\varepsilon_{x,inj}$ = 120 nm and $\varepsilon_{y,inj}$ = 5 nm, injected with optimal beta functions to match the available phase space at the injection point [27]. The injection efficiency can be improved by altering the booster beam parameters, e.g. by having 100% coupling and by reducing the bunch length using a higher RF voltage.

*4.4 Tune working point scans*
After a solution with good Touschek lifetime and acceptable DA was obtained from the above optimisations, a tune working point scan was performed over several units. The tune was changed, while maintaining a fixed chromaticity of (2, 2) and the optimal Twiss functions values found during the optimisation process on all cells (C1c, C2, CINJ). The set of random errors described above was applied and corrected using the techniques quoted above, at each point. Following [49-50], the scan has been performed for several seeds of errors averaging the final results. Fig. 17 shows the variation of DA and lifetime in the tune-space. In this working point scan two regions where both DA and lifetime are simultaneously optimised can be observed: (57.4, 20.1) and (57.4, 21.1). These working points have been tested but did not prove to be as good as the original one when faced to a larger error statistic, leading to several cases with small DA and lifetime.

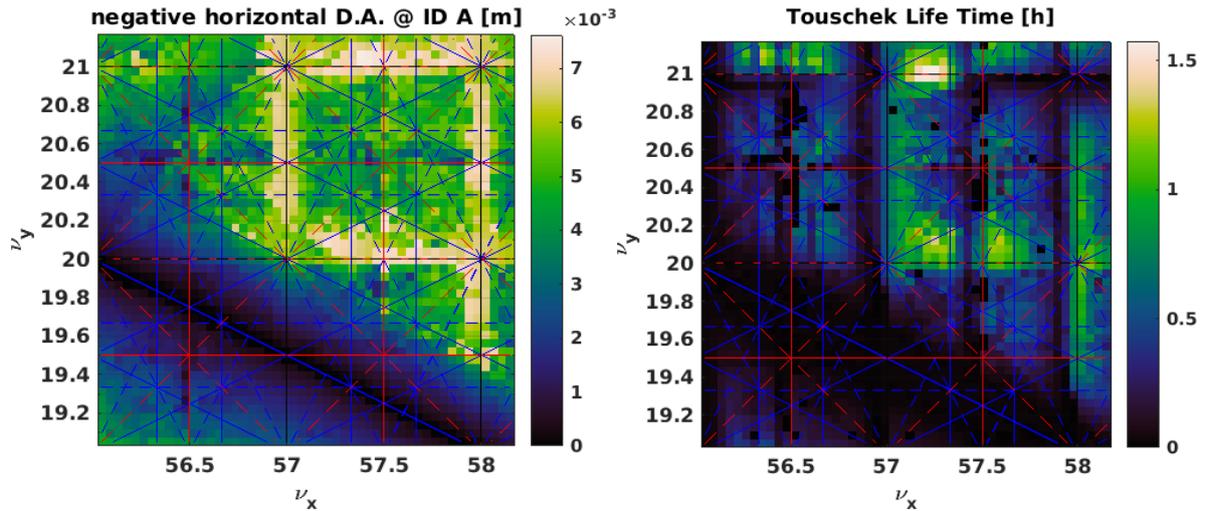

Fig. 17: Tune scans with respect to the DA (left) and Touschek lifetime (right) averaged over 4 random error seeds.

*Conclusion and Acknowledgements*

The Double Triple Bend Achromat (DTBA) is a novel lattice designed for next generation 3 GeV Light Sources. Originating from a modification of the ESRF HMBA cell and inspired by the DDBA lattice studied for Diamond, the DTBA combines the positive characteristics of both designs. A straight at the centre of its cell increases the available drift space for insertion devices; the longitudinal gradient dipoles, together with the combined function dipoles, help to achieve an emittance of 132 pm. Finally the **-I** transformation between the sextupoles, located at two dispersion bumps, enhances the DA and lifetime.

With the extensive linear and non-linear optics optimisations performed on DTBA with no errors, the DA of the lattice is more than ±11.5 mm when a special injection cell is included (DA calculated at ID_D), and the lifetime is τ ~ 2.3 h (zero current bunch length). The performance of the lattice was also studied in the presence of errors and correction: the DA, although decreased to ±10 ±1 mm, is still sufficient for injection; the lifetime is τ = 1.4 ± 0.2h and the injection efficiency 88.3 ± 5%. These values are pessimistic as they assume no bunch elongation due to impedance effects, lower frequency RF systems or harmonic cavities and an unmodified injection chain. With 48 straight sections for insertion devices and 132 pm emittance, the DTBA Light Source promises an unprecedented performance.